\documentclass[conference]{IEEEtran}
\IEEEoverridecommandlockouts
\usepackage[bottom=1.02in, top=0.752in,left=0.65in,right=0.65in]{geometry}
\usepackage{cite}
\usepackage{amsmath,amssymb,amsfonts}
\usepackage[ruled,linesnumbered]{algorithm2e}
\usepackage{graphicx}
\usepackage{textcomp}
\usepackage{xcolor}
\usepackage{multirow}
\usepackage{soul}
\usepackage{hyperref}
\usepackage{bm}
\usepackage[font=small]{caption}
\usepackage{subcaption}

\IEEEaftertitletext{\vspace{-1\baselineskip}} 
\def\BibTeX{{\rm B\kern-.05em{\sc i\kern-.025em b}\kern-.08em
    T\kern-.1667em\lower.7ex\hbox{E}\kern-.125emX}}
\begin{document}

\title{A Scalable Graph Neural Network Decoder for Short Block Codes\\
\thanks{This work has been supported by the SmartSat CRC, whose activities are funded by the Australian Government’s CRC Program.}
}

\author{\IEEEauthorblockN{Kou Tian, Chentao Yue, Changyang She, Yonghui Li, and Branka Vucetic}
\IEEEauthorblockA{School of Electrical and Information Engineering, The University of Sydney, Australia}

}

\maketitle

\begin{abstract}
In this work, we propose a novel decoding algorithm for short block codes based on an edge-weighted graph neural network (EW-GNN).
The EW-GNN decoder operates on the Tanner graph with an iterative message-passing structure, {which algorithmically aligns with the conventional belief propagation (BP) decoding method.}
In each iteration, the ``weight'' on the message passed along each edge is obtained from a fully connected neural network that has the reliability information from nodes/edges as its input. 
Compared to existing deep-learning-based decoding schemes, the EW-GNN decoder is characterised by its scalability, meaning that 1) the number of trainable parameters is independent of the codeword length, and 2) an EW-GNN decoder trained with shorter/simple codes can be directly used for longer/sophisticated codes of different code rates. 
Furthermore, simulation results show that the EW-GNN decoder outperforms {the BP and deep-learning-based BP methods from the literature} in terms of the decoding error rate.
\end{abstract}


\vspace{-0.4em}
\section{Introduction} \label{intro}
\vspace{-0.2em}

Short codes with strong error-correction capability should satisfy the stringent latency and reliability requirements of ultra-reliable low-latency communications (URLLC) in 5G and future 6G networks \cite{shirvanimoghaddam2018short, yue2022efficient}.
Several short linear block codes, including Bose-Chaudhuri-Hocquenghem (BCH) codes and low density parity check (LDPC) codes, are promising candidates for URLLC due to their superior performance in the short block length regions \cite{shirvanimoghaddam2018short}. Although long LDPC codes with the BP decoder have been adopted in the data channel of 5G new radio \cite{3gpp_nrcoding}, their short variants suffer from a significant gap to the finite block length bound \cite{polyanskiy2010channel}. The suboptimal nature of short LDPC codes also results from the limitation of the BP decoder, which has a notable performance degradation compared to the maximum-likelihood (ML) decoding, due to insufficient sparsity in the parity-check matrices \cite{shirvanimoghaddam2018short}. 

In recent years, deep learning (DL) technologies have been applied in the field of channel coding, particularly, the decoder design \cite{deepcoding, nbp2016, nbp2018, hyperbp, buchberger2020pruning, Nachmani2022activation}. In \cite{nbp2016} and \cite{nbp2018}, the authors proposed a neural decoder, namely, the neural BP (NBP), by unrolling the iterative structure of BP to a non-fully connected network with trainable weights. NBP outperforms the conventional BP in terms of decoding error rate when using short BCH codes. 
The NBP decoder is further improved in \cite{hyperbp}, where a hypernetwork structure was developed. On the other hand, the performance of NBP can be improved by optimizing the input parity-check matrices. 
A pruning-based NBP was presented in \cite{buchberger2020pruning}, which removes less-weighted neurons in the NBP network to obtain different parity-check matrices in each iteration.
In addition, the authors of \cite{Nachmani2022activation} proposed efficient and well-suited loss functions to improve the training process of NBP. 

Although aforementioned DL-aided decoding schemes show significantly improved error-correcting performance compared to the conventional BP, several issues remain open. 1) Existing schemes suffer from high complexity, which increases with code length linearly \cite{nbp2016,nbp2018, Nachmani2022activation} or exponentially \cite{deepcoding}. 
2) They are not scalable to the code length and code rate, i.e., we need to re-train the neural networks from scratch when code length or code rate changes. 3) The performance of existing schemes highly depends on the hyper-parameters of neural networks, which are obtained by trail-and-error.

To address the above issues, we leverage the emerging graph neural network (GNN) \cite{gnnsurvey} to develop a novel {edge-weighted GNN (EW-GNN) decoder for short codes}. To simplify the selection of hyper-parameters, EW-GNN aligns with BP \cite{aa}, which operates over Tanner graphs with variable nodes and parity-check nodes. Specifically, each variable node in EW-GNN is assigned a hidden embedding, which is iteratively updated based on the edge messages on the Tanner graph. Then, each edge messages is assigned a ``weight'' determined by a fully-connected feed-forward neural network (FNN).
In each decoding iteration, the FNN updates the weights based on its input features, i.e., reliability information of nodes/edges \cite{LLRbp, info_dec_ldpc}. 
By reducing the weights of unreliable edge messages generated from small cycles in the Tanner graph, EW-GNN can improve the accuracy of the output codeword estimate.
The number of trainable parameters of EW-GNN does not change with the code length. Hence, after the training with a given code, the EW-GNN decoder can be applied to other codes with different rates and lengths.
Simulation results show that EW-GNN outperforms the conventional BP and the NBP \cite{nbp2018} in terms of error rate performance, even when the code length in the training stage is different from that in the testing stage.



\emph{Notation}: We use $a$, $\mathbf{a}$, and $\mathbf{A}$ to represent a scalar, a row vector, and a matrix, respectively. Let $a_i$ denote the $i$-th element of the vector $\mathbf{a}$, and $A_{i,j}$ denote the element at the $i$-th row and the $j$-th column of the matrix $\mathbf{A}$.
We use $(\cdot)^\intercal$ to represent the transpose of a vector or a matrix, e.g., $\mathbf{a}^\intercal$ and $\mathbf{A}^\intercal$. The calligraphic letters denote sets, e.g., the set $\mathcal{A}$. In particular, $\mathcal{N}(\mu, \sigma^2)$ stands for the normal distribution and $\mathbb{R}$ is the set of real numbers. $\mathrm{Pr}(\cdot)$ 
denotes the probability of an event, and $|\cdot|$ is the operator of absolute value.
\vspace{-0.4em}
\section{Preliminaries}
\vspace{-0.3em}
\subsection{Linear Block Codes and Tanner Graph} \label{Sec::Preliminaries::TG}
\vspace{-0.3em}
We consider a binary linear block code  $ \mathcal{C}(n,k) $ that is defined by a generator matrix $\mathbf{G} \in \{0,1\}^{k \times n}$ or a parity-check matrix $\mathbf{H} \in \{0,1\}^{(n-k) \times n}$, where $n$ and $k$ denote the codeword length and message length, respectively. 
For a binary message sequence $\mathbf{b}\in \{0,1\}^k$, its corresponding 
codeword $\mathbf{c}$ is generated by the encoding operation  $\mathbf{c} = \mathbf{b}\mathbf{G}$ and satisfies  $\mathbf{H}\mathbf{c}^\intercal = \mathbf{0}$. 
We assume that the modulated symbol $\mathbf{x}$ of $\mathbf{c}$ is transmitted over the additive white Gaussian noise (AWGN) channel with the binary phase shift keying (BPSK) modulation. Thus,  $\mathbf{x}$ is given by $\mathbf{x} = 1 - 2{\mathbf{c}}\in \{-1, 1\}^n$, and the noisy received signal $\mathbf{y}$ is obtained as $\mathbf{y} = \mathbf{x} + \mathbf{z}$, where $\mathbf{z}$ is the AWGN vector with noise power $\sigma^2$, i.e., $z_i  \sim \mathcal{N}(0, \sigma^2) $. We define the signal-to-noise ratio (SNR) in dB as $\gamma= 10\log_{10}(1/\sigma^2)$. 
At the receiver side, a soft-decision decoder decodes $\mathbf{y}$ and outputs an estimate $\hat{\mathbf{c}}$ of the transmitted codeword $\mathbf{c}$. A decoding error occurs when $\hat{\mathbf{c}} \neq \mathbf{c}$.

\begin{figure}[tb]
\centering
\includegraphics[scale=0.7]{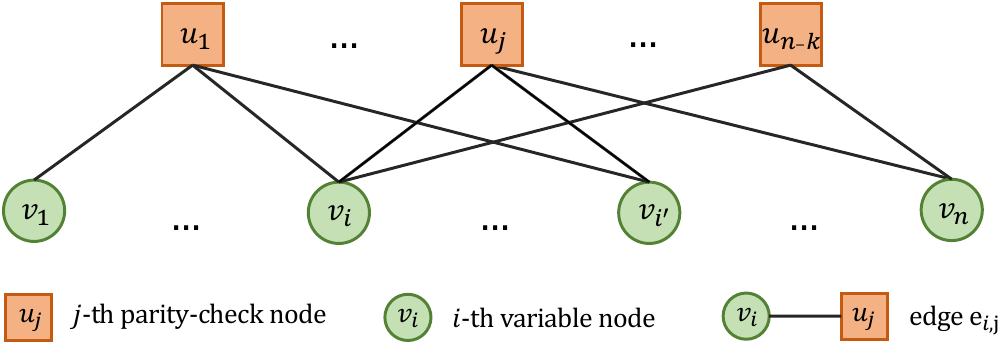}
\caption{A bipartite Tanner graph $\mathcal{G}=(\mathcal{V}, \mathcal{U}, \mathcal{E})$.}
\vspace{-0.2em}
\label{fig::tannergraph}
\end{figure}

Tanner graph is a graphical representation of linear block codes \cite{tanner}. As illustrated in Fig. \ref{fig::tannergraph}, for a code $ \mathcal{C}(n,k) $ with the parity-check matrix $\mathbf{H}$, its Tanner graph consists of two sets of nodes: $n$ variable nodes corresponding to codeword bits and $n-k$ check nodes corresponding to parity-check equations. The $i$-th variable node $v_i$ is connected to the $j$-th check node $u_j$ if the $(j,i)$-th element of $\mathbf{H}$ is one. In this case, H can be interpreted as the adjacency matrix of a bipartite graph $ \mathcal{G}=(\mathcal{V}, \mathcal{U}, \mathcal{E})$ with two disjoint node sets $\mathcal{V} = \{v_1,...,v_{n}\}$ and $\mathcal{U} = \{u_1,...,u_{n-k}\}$, and an edge set defined as $\mathcal{E} = \{e_{i,j} = (v_i, u_j): v_i\in\mathcal{V},u_j\in\mathcal{U}, H_{j,i}=1\} $.  
We denote the neighborhood of a variable node $v_i$, i.e., all check nodes connected to $v_i$, as  $\mathcal{M}(v_i) = \{u_j \in \mathcal{U} : e_{i,j} \!\in \!\mathcal{E}\}$. Similarly, $\mathcal{M}(u_j) = \{v_i \in \mathcal{U} : e_{i,j} \!\in\! \mathcal{E}\}$ represents the set of neighbor variable nodes of a check node $u_j$.

\vspace{-0.3em}
\subsection{Iterative Decoding Algorithm}
\vspace{-0.3em}
BP is a 
soft-decision decoding algorithm that accepts soft information (i.e., the bit-wise log-likelihood ratios (LLRs) observed from channel) as its input \cite{LLRbp}. We denote the channel LLR sequence of codeword $\mathbf{c}$ as $\mathbf{s} \in \mathbb{R}^n$. For the BPSK modulation over an AWGN channel, $s_i$ can be obtained by 
\begin{equation} \label{equ::llr}
s_i = \ln{\frac{\mathrm{Pr}(c_i = 0|y_i)}{\mathrm{Pr}(c_i = 1|y_i)}} = \frac{2y_i}{\sigma^2}.
\end{equation}


The BP decoder performs iterative message-passing on the Tanner graph, 
by transmitting messages between variable nodes and check nodes. Specifically, at the {$t$-th iteration}, the message passed from $u_j$ to $v_i$ is given by 
\begin{equation} \label{equ::u_to_v}
    \mu_{u_j \to v_i}^{(t)} = 2\tanh^{-1}\left(\prod_{v \in \mathcal{M}(u_j)\setminus v_i} \tanh \left(\frac{\mu_{v \to u_j}^{(t-1)}}{2} \right) \right),
\end{equation}
for each $u_j \in \mathcal{U}$ and $v_i \in \mathcal{M}(u_j)$. 
The message passed from $v_i$ to $u_j$ is given by
\begin{equation}  \label{equ::v_to_u}
    \mu_{v_i \to u_j}^{(t)}= s_i + \sum_{u \in \mathcal{M}(v_i)\setminus u_j}\mu_{u \to v_i}^{(t)},
\end{equation}
for each $v_i \in \mathcal{V}$ and $u_j \in \mathcal{M}(v_i)$.


At the beginning of decoding (i.e., $t=1$), the message $\mu_{v_i \to u}^{(0)}$ is initialized by the channel LLR corresponding to variable node $v_i$, i.e., $\mu_{v_i \to u}^{(0)} = s_i$ for $v_i\in\mathcal{V}$ and $u\in\mathcal{M}(v_i)$. Then, BP iteratively performs \eqref{equ::u_to_v} and \eqref{equ::v_to_u}, and updates the messages $\mu_{u_j \to v_i}^{(t)}$ and $\mu_{v_i \to u_j}^{(t)}$ along edges in $\mathcal{E}$. This iterative process continues until $\mu_{u_j \to v_i}^{(t)}$ and $\mu_{v_i \to u_j}^{(t)}$ converge (e.g., see the analysis in \cite{richardson2001capacity}), or the allowed maximum number of iterations is reached. Finally, assuming that $T$ is the last iteration, BP obtains the soft information of $v_i\in\mathcal{V}$ as

\vspace{-0.3em}
\begin{equation} \label{equ::BPoutput}
    \ell_{v_i} = s_i + \sum_{u \in \mathcal{M}(v_i)}\mu_{u \to v_i}^{(T)},
\end{equation}
\vspace{-0.3em}
Here, $\ell_{v_i}$ is known as the \textit{a posteriori} LLR 
of the $i$-th codeword bit $c_i$.
Therefore, we can obtain $\hat{c}_i$, which is the estimate of the transmitted bit $c_i$, based on $\ell_{v_i}$. Specifically, $\hat{c}_i= 0$ if $\ell_{v_i}>0$, and $\hat{c}_i= 1$ otherwise.

The decoding performance of BP depends on the structure of codes (i.e., structure of Tanner graphs). To be more precise, it is strongly limited by short \textit{cycles} existing in the Tanner graph. 
According to \cite{fgandspa}, small cycles introduce correlations between $\mu_{u \to v}^{(t)}$ and $\mu_{v \to u}^{(t')}$ across decoding iterations where $t\neq t'$, thereby preventing the posterior LLR $\ell_{v_i}$ being exact. For example, for BCH codes and short LDPC codes, BP has inferior performance due to the small cycles of their Tanner graphs \cite{multi-baseBP, shirvanimoghaddam2018short}. 





\vspace{-0.3em}
\section{Edge-Weighted GNN Decoding Algorithm}

\vspace{-0.3em}
\subsection{Graph Neural Networks and Algorithmic Alignment}
\vspace{-0.3em}
GNN is developed for processing graph structured data.
It learns the low-dimensional representation of graphs, i.e., the network \textit{embedding}, by iteratively propagating and aggregating the node/edge information over the graph \cite{gnnsurvey}. 
Embeddings capture the graph topology, and they are used for graph analysis tasks, e.g., node/graph classification and link prediction. 
We note that the channel decoding problem is in fact a node classification task on the Tanner graph, as demonstrated in Section \ref{Sec::Preliminaries::TG}.

Generally, many existing GNNs can be represented as a message passing neural network (MPNN)\cite{mpnn}. 
For a general input graph with node features, MPNN generates node embeddings with two differentiable functions, namely, the update function $U_t(\cdot)$ and the message function $M_t(\cdot)$. 
For each edge in the graph, $M_t(\cdot)$ computes a ``message'' based on the node embeddings of its endpoints. Then, for each node, $U_t(\cdot)$ updates the embedding value by aggregating messages from its neighbors. For details, we refer readers to \cite{mpnn}


GNN is capable of learning well a wide range of practical models based on graphs with superior performance compared to the {traditional  FNN}.
This is justified by the \textit{algorithmic alignment} framework \cite{aa}.
Generally speaking, a neural network aligns well with a classical algorithm if the algorithm can be partitioned into multiple parts, each of which can be conveniently modeled by one of the modules in the neural network\footnote{We refer interested readers to \cite[Definition 3.4]{aa} for its formal definition.}. 
GNN can align with (in other words, learn well) the BP decoding algorithm,  because 1) BP solves node classification problems on the Tanner graph and 2)  both GNN and Tanner graph are permutation-invariant (due to the linear property of codes) \cite[Section 2]{aa}. 


\begin{figure}[t]
\centering
\includegraphics[scale=0.95]{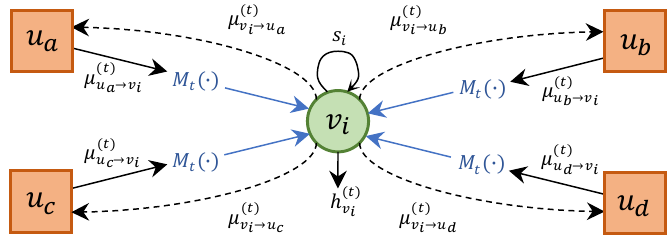}
\caption{Illustration of updating the node embedding in the GNN-based decoder. The variable node $v_i$ receives messages from $\mathcal{M}(v_i)=\{u_a, u_b, u_c, u_d\}$.}
\vspace{-0.2em}
\label{fig::emb}
\end{figure}

Motivated by the algorithmic alignment framework of GNN, we propose a GNN-based scalable decoder for linear block codes.
Considering a bipartite graph $ \mathcal{G}=(\mathcal{V}, \mathcal{U}, \mathcal{E})$ with directed edge messages, as introduced in Section \ref{Sec::Preliminaries::TG}, 
we define the node feature of variable node $v_i$ as the received LLR information\footnote{GNN are capable of processing various types of nodes by applying different update and message functions.
In the GNN decoder, the node classification only performs on variable nodes that are associated with codeword bits. Thus, we only assign features and embeddings to variable nodes for simplicity.}, $s_i$ given by \eqref{equ::llr}.
In each iteration, the hidden node embedding of $v_i$ is updated based on incoming edge messages $\mu_{u \to v_i}$ according to
\begin{equation} \label{equ::gnn_decoder}
    h_{v_i}^{(t)}=U_t\Big(s_i, \sum_{u \in \mathcal{M}(v_i)} M_t(\mu_{u \to v_i}^{(t)},h_{v_i}^{(t-1)})\Big),
\end{equation}
where the edge message $\mu_{u \to v}^{(t)}$ is iteratively generated from $\mu_{v \to u}^{(t-1)}$. Fig. \ref{fig::emb} gives an example of the updating process in \eqref{equ::gnn_decoder}, which will be further elaborated in Section \ref{sec::EW-GNN}. Since $n$ variable nodes correspond to $n$ codeword bits, the GNN decoder finally obtains ${\hat{c}_{i}}$ from $h_{v_i}^{(T)}$, for $v_i \in \mathcal{V}$, {after} $T$ {iterations}.
\vspace{-0.3em}
\subsection{GNN Decoder with Edge Weights}  \label{sec::EW-GNN}
\vspace{-0.3em}

In the proposed GNN decoder, we introduce ``weight'' to the edge messages to improve the decoding performance.
{Intuitively, by reducing the weights of} unreliable edge messages
generated from small cycles in the Tanner graph, {it is possible to} improve the  accuracy of the output codeword estimate $\hat{\mathbf{c}}$. 

The weight is a multiplicative correction factor \cite{yazdani2004cycleBP}  working on the edge message from parity-check nodes to variable nodes.
Let us denote the weight of $\mu_{u_j \to v_i}^{(t)}$ in the $t$-th iteration as $w_{u_j \to v_i}^{(t)} \in \mathbb{R}$.
By employing {an FNN}, $g_{\bm{\theta}}(\cdot)$ with trainable parameters $\bm{\theta}$, 
the weight $w_{u_j \to v_i}^{(t)}$ is learned from input features 
that can reflect the reliability of node and edge information, which is defined as
\begin{equation} \label{equ::ewgnn_weight}
\begin{split}
    w_{u_j \!\to v_i}^{(t)} \!\!=\! g_{\bm{\theta}}(|\mu_{u_j \!\to v_i}^{(t)}\!|,  r^{\!(t)}(\mu_{u_j \!\to v_i}\!),
    r^{\!(t\!-\!1)}(\mu_{v_i \!\to u_j}\!), 
    r^{\!(t\!-\!1)}({h}_{v_i}\!)).
\end{split}
\end{equation}
In \eqref{equ::ewgnn_weight}, $|\mu_{u_j \to v_i}^{(t)}|$ represents the reliability of edge message\footnote{This is because $\mu_{u_j \to v_i}^{(t)}$ can be regarded as an extrinsic LLR (see \cite{LLRbp})} $\mu_{u_j \to v_i}^{(t)}$. 
Then, $r^{(t)}(x)$ is a residual function given by
\begin{equation} \label{equ:ressidual}
    r^{(t)}(x) = |x^{(t)}-x^{(t-1)}|.
\end{equation}
Thus, $r^{(t)}(\mu_{u_j \to v_i})$, $ r^{(t-1)}(\mu_{v_i \to u_j})$, and $r^{(t-1)}({h}_{v_i})$ represent the residual value of $\mu_{u_j \to v_i}$, $\mu_{v_i \to u_j}$, and ${h}_{v_i}$, respectively.
They are also measurements of the reliabilities \cite{info_dec_ldpc}; precisely, a lower residual value is, 
a more reliable node/edge will be.


Next, we explain the algorithm of EW-GNN with the help of weight $w_{u_j \to v_i}^{(t)}$.
At the $t$-th iteration, EW-GNN first updates the edge message from $u_j$ to $v_i$ according to
\begin{equation} \label{equ::ewgnn_utov}
     \mu_{u_j \to v_i}^{(t)} = \ln 
     \frac{f_{c}\left(
     1+ 
     \prod_{v \in \mathcal{M}(u_j)\setminus v_i} 
     \tanh \left(
     \frac{\mu_{v \to u_j}^{(t-1)}}{2}
     \right),
     \alpha \right)}
     {f_{c}\left(
     1- 
     \prod_{v \in \mathcal{M}(u_j)\setminus v_i} 
     \tanh \left(
     \frac{\mu_{v \to u_j}^{(t-1)}}{2} 
     \right),
     \alpha \right)}
     ,
\end{equation}
where
\begin{equation}
    f_c(x, \alpha) = 
    \begin{cases}
    2-\alpha, &  x > 2 - \alpha,\\
    x, &  \alpha \leq x \leq 2-\alpha, \\
    \alpha, & x < \alpha, \\
    \end{cases}
\end{equation}
and $\alpha$ is a model parameter less than $10^{-7}$.
We note that the message update function \eqref{equ::u_to_v} in BP is {not applicable} to neural-network-based decoders, because $\tanh^{-1}(\cdot)$ {is unbounded} and introduces non-differentiable singularities, leading to a non-convergence training process. Therefore, {we modify} $\tanh^{-1}(\cdot)$ with the clip function $f_{c}(\cdot)$ 
to avoid the aforementioned issue. 
Compared to the Taylor series expansion of $\tanh^{-1}(\cdot)$ used in \cite{hyperbp}, $f_{c}(\cdot)$ in \eqref{equ::ewgnn_utov} introduces virtually no extra complexity. 

After $\mu_{u_j \to v_i}^{(t)}$ is obtained for $u_j \in \mathcal{U}$ and $v_i \in \mathcal{M}(u_j)$, the updated edge message from $v_i$ to $u_j$ is given by\footnote{$M_t\!(\mu_{u_j \!\to v_i}^{(t)}\!, h_{v_i}^{(t\!-\!1)}) \!=\! w_{u_j \!\to v_i}^{(t)} \mu_{u_j \!\to v_i}^{(t)}$ as $w_{u_j \!\to v_i}^{(t)}$ is a function of $h_{v_i}^{(t\!-\!1)}\!\!.$}
\begin{equation} \label{equ::ewgnn_vtou}
\begin{split}
    \mu_{v_i \to u_j}^{(t)}&=  s_i + \sum_{u \in \mathcal{M}(v_i)\setminus u_j}M_t(\mu_{u_j \to v_i}^{(t)},  h_{v_i}^{(t-1)})\\
    &= s_i + \sum_{u \in \mathcal{M}(v_i)\setminus u_j}w_{{u \to v_i}}^{(t)} \cdot \mu_{u \to v_i}^{(t)}.
\end{split}
\end{equation}
Then, the update function $U_t(\cdot)$ {aggregates} weighted edge messages from all neighbor nodes, and {updates} node embedding  $h_v^{(t)}$ {according to}
\begin{equation} \label{equ::ewgnn_emb}
\begin{split}
     {h}_{v_i}^{(t)}&=U_t\Big(s_i, \sum_{u \in \mathcal{M}(v_i)} M_t(\mu_{u \to v_i}^{(t)},h_{v_i}^{(t-1)}\Big),\\
     &= s_i + \sum_{u \in \mathcal{M}(v_i)}w_{{u \to v_i}}^{(t)} \cdot \mu_{u \to v_i}^{(t)}. 
\end{split}
\end{equation}

\begin{figure}[t]
\centering
\includegraphics[scale=0.5]{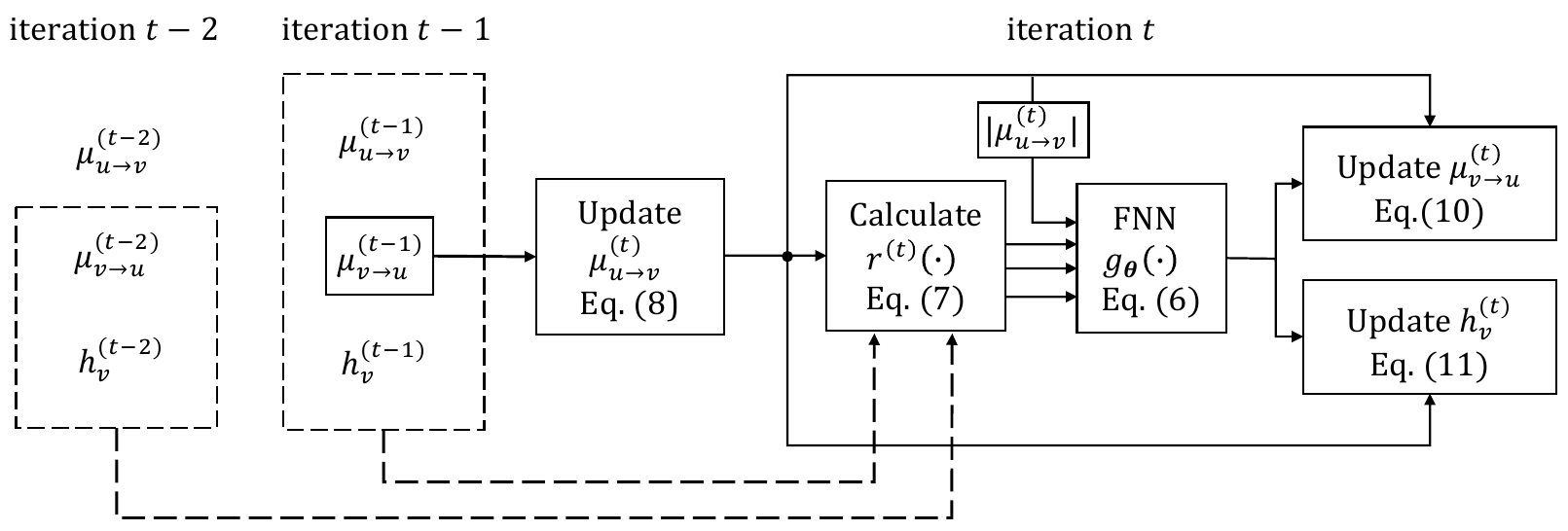}
\caption{The information flow across adjacent iterations in the EW-GNN.}
\label{fig::info_flow}
\vspace{-0.2em}
\end{figure}
At $t=0$, the node embeddings $h_{v_i}^{(0)}$ and edge messages $\mu_{v_i \to u}^{(0)}$ are all initialized to corresponding input features of node $v_i$, i.e., $s_i$ given in \eqref{equ::llr}. Also, all residuals are initialized to 0.
{Meanwhile}, at the $t$-th iteration, EW-GNN uses node embeddings and edge messages propagated from previous two iterations to update $h_v^{(t)}$, as demonstrated in Fig. \ref{fig::info_flow}.
After $T$ iterations, we obtain the estimated codeword $\hat{\mathbf{c}}$ from the node embedding ${h}_{v_i}$ according to the following rule:
\begin{equation} \label{equ::ewgnn_readout}
    \hat{c}_i = 
    \begin{cases}
    1, & {h}^{(T)}_{v_i} \leq 0,\\
    0, & {h}^{(T)}_{v_i} > 0.\\
    \end{cases}
\end{equation}

The algorithm of the proposed EW-GNN decoder is summarized in Algorithm \ref{alg::ewgnn}, which includes the training phase and {the} inference phase. Thanks to the scalability of EW-GNN decoder, one can apply different Tanner graphs (i.e.,  $\mathbf{H}\neq \mathbf{H}^\prime$ and $\mathcal{G}\neq \mathcal{G}^\prime$) in training and inference phases; this will be further demonstrated in Section \ref{Sec::experiments}.

\begin{algorithm} [t]
\small 
\caption{Training and Inference of EW-GNN}
\label{alg::ewgnn}
\SetKwComment{Comment}{$\triangleright$\ }{}
\SetKwInOut{Input}{Input}\SetKwInOut{Output}{Output}
\SetKwFor{For}{for}{do}{endfor} 
\Comment{End-to-end Training Phase:}
\Input{Generator matrix $\mathbf{G}$, parity-check matrix $\mathbf{H}$, training batch size $N_B$, training SNR $[\,\gamma_{\min}, \gamma_{\max}]\,$, number of iterations $T$, learning rate $\eta$ \\}
\Output{well-trained $g_{\bm{\theta}^{*}}(\cdot)$}
Create the network topology, $ \mathcal{G}=(\mathcal{V}, \mathcal{U}, \mathcal{E})$ from $\mathbf{H}$;\\
\ForEach{epoch}{
    \For{$n_b \leftarrow 1 : N_B $}{
        Obtain codeword $\mathbf{c} = \mathbf{b} \mathbf{G} $ from a random message $\mathbf{b}$;\\
        Obtain the noisy signal $\mathbf{y}$ with the AWGN channel at a random SNR $\gamma \in [\,\gamma_{\min},\gamma_{\max}]\,$;\\
        Obtain the received LLR vector $\mathbf{s}$ by \eqref{equ::llr};\\
        Initialize $h_{v_i}^{(0)}\!=\! s_i$, $\mu_{v_i \!\to u}^{(0)} \!=\! s_i$, $\mu_{u \!\to v} ^{(0)} \!=\! 0$, $r^{(0)}\!(\cdot) \!=\! 0$;\\
        \For{$t \leftarrow 1 : T$}{
            \ForEach{$e_{i,j} \in \mathcal{E}$}{
            Update edge message $\mu_{u_j \to v_i}^{(t)}$ by \eqref{equ::ewgnn_utov};\\
            Update weight $w_{u_j \to v_i}^{(t)}$ by \eqref{equ::ewgnn_weight};\\
            Update edge message $\mu_{v_i \to u_j}^{(t)}$ by \eqref{equ::ewgnn_vtou};\\
            }
            \ForEach{$v_i \in \mathcal{V}$}{
            Update node embedding $h_{v_i}^{(t)}$ by \eqref{equ::ewgnn_emb};
            }
        }
        Calculate the multiloss value $L_{n_b}$ by \eqref{equ::lossfuncion};
    }
    Update $\bm{\theta}$ with the loss  $L(\bm{\theta})=\frac{1}{N_B}\sum_{n_b=1}^{N_B}L_{n_b}$ by the Adam optimizer with learning rate $\eta$;
}
\BlankLine
\Comment{Inference Phase:}
\Input{LLRs of the received codewords $\mathbf{s}\in \mathbb{R}^n$, parity-check matrix $\mathbf{H^\prime}$, number of iterations $T^\prime$}
\Output{estimated codewords $\hat{\mathbf{c}}$}
Create the network topology, $ \mathcal{G}^\prime=(\mathcal{V}^\prime, \mathcal{U}^\prime, \mathcal{E}^\prime)$ from $\mathbf{H}^\prime$;\\
Execute line 7 to line 17 on $ \mathcal{G}^\prime$  with  $g_{\bm{\theta^{*}}}(\cdot)$ for $T'$ iterations;\\
Obtain the estimated codewords $\hat{\mathbf{c}}$ according to \eqref{equ::ewgnn_readout}.
\vspace{-0.1em}
\end{algorithm}

\vspace{-0.3em}
\subsection{Training {Algorithm}}
\vspace{-0.3em}

The goal of the training phase is to find an optimal FNN $g_{\bm{\theta}^{*}}(\cdot)$  that minimizes the difference between the transmitted codeword and the estimate output by the EW-GNN decoder. 
For channel coding, a large amount of data samples can be easily obtained to perform the end-to-end training  \cite{deepcoding}.
The label data, i.e., the transmitted codewords, are generated by encoding a random binary message. 
By imposing an AWGN with power $\sigma^2$ over the symbols of codewords, received LLRs can be computed according to \eqref{equ::llr}, which are used as the input data of EW-GNN. To improve the robustness of EW-GNN, we perform the training process with diverse SNRs uniformly distributed in the interval $[\gamma_{\min}, \gamma_{\max}]\,$. 

Since the channel decoding can be considered as a bit-wise binary classification task, we apply the average binary cross-entropy (BCE) per bit per iteration as the loss function, i.e., 
\begin{equation} \label{equ::lossfuncion}
    L = -\frac{1}{n T}\sum_{t=1}^{T}\sum_{i=1}^{n}{c_i \log(p^{(t)}_{v_i})+(1-c_i) \log(1-p^{(t)}_{v_i})},
\end{equation}
which is the multiloss variant \cite{nbp2018} of a regular BCE loss function and allows the EW-GNN to learn from early network layers. 
In \eqref{equ::lossfuncion}, $p^{(t)}_{v_i}$ denotes the probability $\mathrm{Pr}(c_i = 1)$ estimated from the node embedding of $v_i$ at iteration $t$, which is given by 
\begin{equation} 
    p^{(t)}_{v_i}=\frac{1}{1+e^{{h}_{v_i}^{(t)}}}\in (0,1).
\end{equation}
During the training phase, FNN parameters are updated utilizing the Adam optimizer with a given learning rate $\eta$ \cite{adam}.

\vspace{-0.5em}
\subsection{Design of FNN}
\vspace{-0.3em}
We determine the normalization technique and activation function used for the FNN by exploiting the domain knowledge in channel coding. 

\subsubsection{Input Normalization} 
According to \eqref{equ::ewgnn_weight} and \eqref{equ:ressidual}, the input features of the FNN includes the absolute value of LLR, which varies widely in $[0, \infty)$. In this case, feature normalization is necessary for better convergence in training. Since all input features of $g_{\bm{\theta}}(\cdot)$ is non-negative, 
they and their transformations cannot be assumed to follow normal distributions. As a result, the standard score normalization is not applicable to EW-GNN,
and we normalize the input feature by its mean value. 
For example, the input feature $|\mu_{u_j \to v_i}^{(t)}|$ is normalized to 
\begin{equation} \label{equ::normalization}
    |\tilde{\mu}_{u_j \to v_i}^{(t)}| = \frac{|\mu_{u_j \to v_i}^{(t)}|}{\frac{1}{E}\sum_{e_{i,j}\in \mathcal{E}}|\mu_{u_j \to v_i}^{(t)}|},
\end{equation}
where $E$ is total number of edges in the graph, given by $E=\sum {H}_{j,i}$. Other input features of $g_{\bm{\theta}}(\cdot)$ are normalized in the same manner as \eqref{equ::normalization}.
\subsubsection{Activation Function} In the EW-GNN, we use the exponential linear unit (ELU) as the activation function for each FNN layer, which is defined as
\begin{equation}
    f_{elu}(x) = 
    \begin{cases}
    x, & x>0,\\
    \beta(e^x-1), &x\leq0,
    \end{cases}
\end{equation}
ELU allows to produce negative output weights with small values. It is found experimentally that ELU performs better than the rectified linear unit (ReLu) activation function which only has non-negative outputs, in training of the EW-GNN decoder.

\vspace{-0.5em}
\subsection{Complexity Analysis}
\vspace{-0.2em}
We analyze the computational complexity of the EW-GNN decoding algorithm, and compare it with those of the conventional 
BP decoder and the DL-based NBP decoder \cite{nbp2018}.

\subsubsection{Inference Complexity}
BP, NBP, and EW-GNN {are} message-passing decoding algorithms. Thus, their overall complexity can be characterized by the number of iterations and the number of operations (e.g., additions and multiplications) per iteration. 

In each BP iteration, \eqref{equ::u_to_v} and \eqref{equ::v_to_u} are in fact computed for each $e_{i,j} \in \mathcal{E}$. Let $\Omega_1$ denote the number of operations in computing \eqref{equ::u_to_v} and \eqref{equ::v_to_u}. Then, the complexity of the BP decoder with $T$ iterations is represented as $O(\Omega_1 E \cdot T)$. Note that \eqref{equ::BPoutput} is readily obtained from \eqref{equ::v_to_u} at iteration $T$; thus, its complexity is negligible. Compared with the BP, a NBP decoder adds extra multiplications operation to \eqref{equ::v_to_u} by assigning $E$ weights to corresponding edge messages. At the final iteration $T$, $E$ additional weights are applied to \eqref{equ::BPoutput} to obtain the decoding results. Therefore, the complexity of the NBP decoder is represented as $O((\Omega_1 +1)E \cdot T + E)$.

In the proposed EW-GNN algorithm, the weight $w_{u_j \to v_i}^{(t)}$ is generated on-the-fly by a trained neural network according to \eqref{equ::ewgnn_weight}. 
Let $\Omega_2$\ denote the total number of operations involved in the forward propagation of $g_{\bm{\theta^{*}}}(\cdot)$. Then, the complexity of the EW-GNN decoder with $T$ iterations is represented as $O((\Omega_1 + \Omega_2 + 1)E \cdot T)$. Here, $\Omega_2$ is dependent on the architecture of the neural network. 

Since $E=\bar{d}_v n$ for $\bar{d}_v$ being the average degree of variable nodes, the complexity of all these algorithms increase linearly with the code length $n$. 

\subsubsection{Training complexity}
We use the number of trainable parameters in neural networks to represent the training complexity. 
Since the NBP decoder assigns in total $2E$ trainable parameters (weights) to edge messages in \eqref{equ::v_to_u} and  \eqref{equ::BPoutput} \cite[Section \uppercase\expandafter{\romannumeral5}]{nbp2018}, it has the training complexity $2E$, which increases linearly with the length of input codewords.

Different from existing DL-aided decoding algorithms in the literature \cite{deepcoding, nbp2016, nbp2018, hyperbp, buchberger2020pruning, Nachmani2022activation}, the training complexity of EW-GNN is independent of the code length and code rate $k/n$, and only depends on the network size of $g_{\bm{\theta}}(\cdot)$. 
In this work, we apply a 3-layer FNN with 32 hidden units for $g_{\bm{\theta}}(\cdot)$, whose training complexity is $\Omega_t= 1249$ for arbitrary block codes. In this case, the proposed EW-GNN has a lower training complexity than the NBP algorithm for large $E$ (e.g., for large code length $n$ and dense parity-check matrix $\mathbf{H}$). 
\vspace{-0.5em}
\section{Experiments and Results} \label{Sec::experiments}
\vspace{-0.5em}

We evaluate the error-correction performance and the scalability of the proposed EW-GNN for short BCH codes and LDPC codes. 
The BP and NBP decoders \cite{nbp2018} are considered as benchmarks for performance comparison. 
The decoding performance of these algorithms are measured by bit error rates (BER) at various SNRs. 
{The} BER is obtained by performing simulations until at least 10,000 bit errors are collected.
The training of the EW-GNN algorithm is implemented using TensorFlow 
with hyper-parameters in Table \ref{tab::parameters}. 
\begin{table}[tb]
\renewcommand\arraystretch{1.2}   
\centering
\caption{Hyper-parameters of the proposed EW-GNN}
\label{tab::parameters}
\begin{tabular}{lcc}
\hline
\multirow{1}{*}{Parameters} 
       &  \multicolumn{1}{c}{BCH codes} & LDPC codes \\ \hline \hline
    Clip factor $\alpha$  & \multicolumn{1}{c}{$10^{-32}$} &  $10^{-7}$ \\ \hline
    SNR range  $[\gamma_{\min}, \gamma_{\max}]$    & \multicolumn{1}{c}{$[3 ~\mathrm{dB}, 8 ~\mathrm{dB}]$} & $[1 ~\mathrm{dB}, 8 ~\mathrm{dB}]$ \\ \hline
    Batch size  $N_B$            & \multicolumn{1}{c}{2000} &  4000 \\ \hline
    Learning rate {$\eta$}   & \multicolumn{2}{c}{$10^{-3}\sim 10^{5}$}    \\ \hline
     Number of iterations $T$            & \multicolumn{2}{c}{8}    \\ \hline
     FNN layers             & \multicolumn{2}{c}{3}    \\ \hline
     Number of hidden units             & \multicolumn{2}{c}{32}    \\ \hline
     ELU factor $\beta$          & \multicolumn{2}{c}{1.0}    \\ \hline
\end{tabular}
\end{table}

\begin{figure*}[bt]
     \centering
     \begin{subfigure}[b]{0.3\textwidth}
         \centering
         \includegraphics[width=\textwidth]{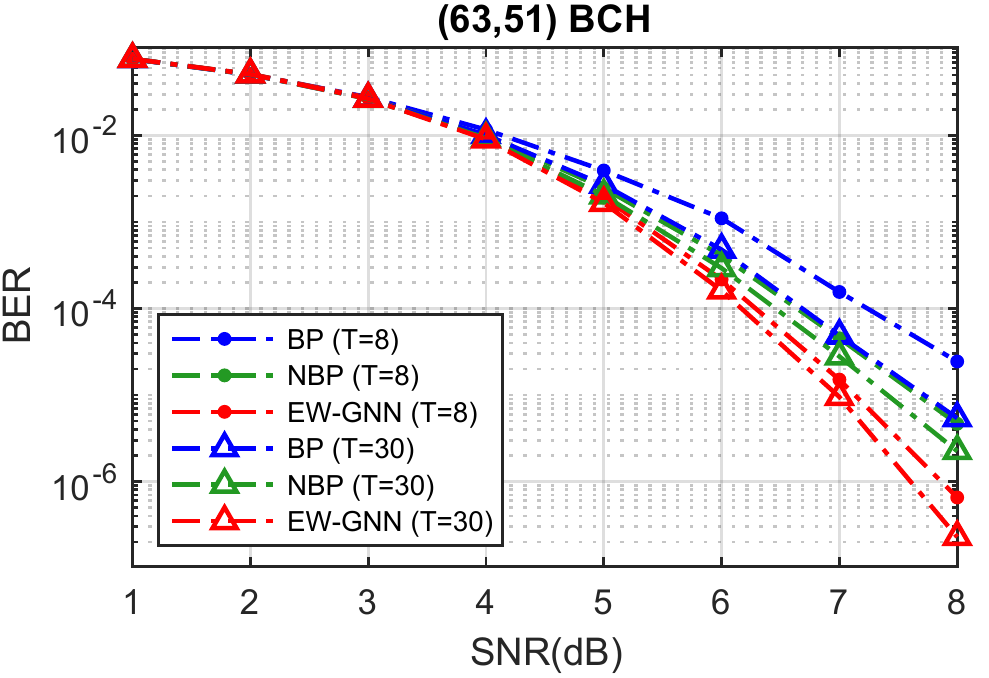}
         \caption{$k=51$}
         \label{fig::6351bch}
     \end{subfigure}
     \hfill
     \begin{subfigure}[b]{0.3\textwidth}
         \centering
         \includegraphics[width=\textwidth]{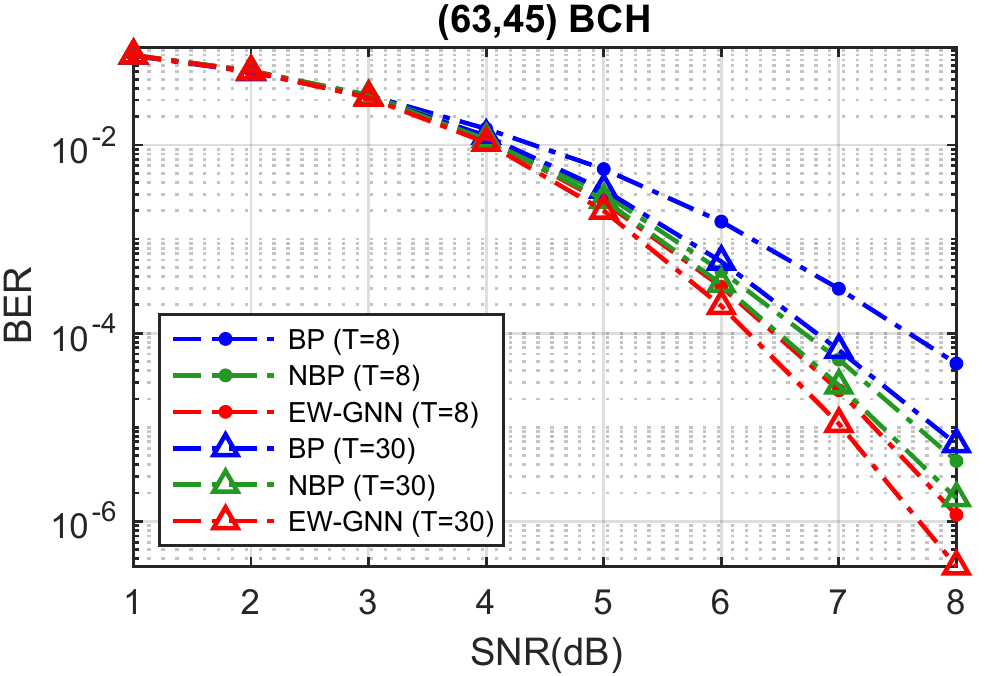}
         \caption{$k=45$}
         \label{fig::6345bch}
     \end{subfigure}
     \hfill
     \begin{subfigure}[b]{0.3\textwidth}
         \centering
         \includegraphics[width=\textwidth]{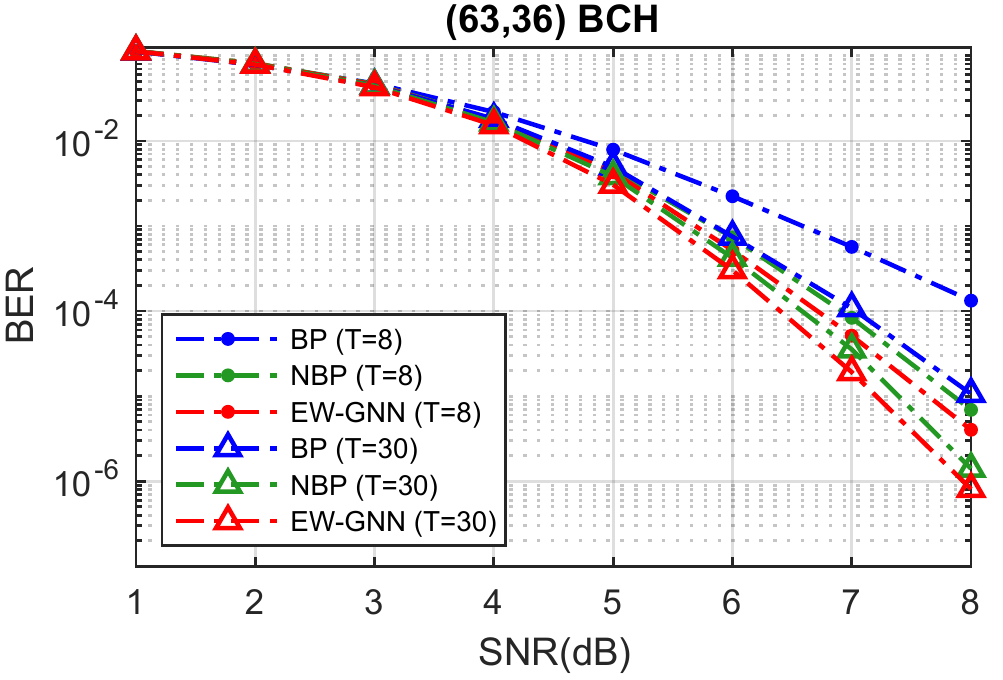}
         \caption{$k=36$}
         \label{fig::6336bch}
     \end{subfigure}
        \vspace{-0.5em}
        \caption{BER performance for BCH codes of length $n=63$. EW-GNN is only trained with the $(63,51)$ BCH code and $T=8$.}
        \vspace{-1em}
        \label{fig::bch_results}
    \centering
\end{figure*}

\begin{figure*}[t]
    \centering
    \begin{subfigure}[b]{0.3\textwidth}
         \centering
         \includegraphics[width=\textwidth]{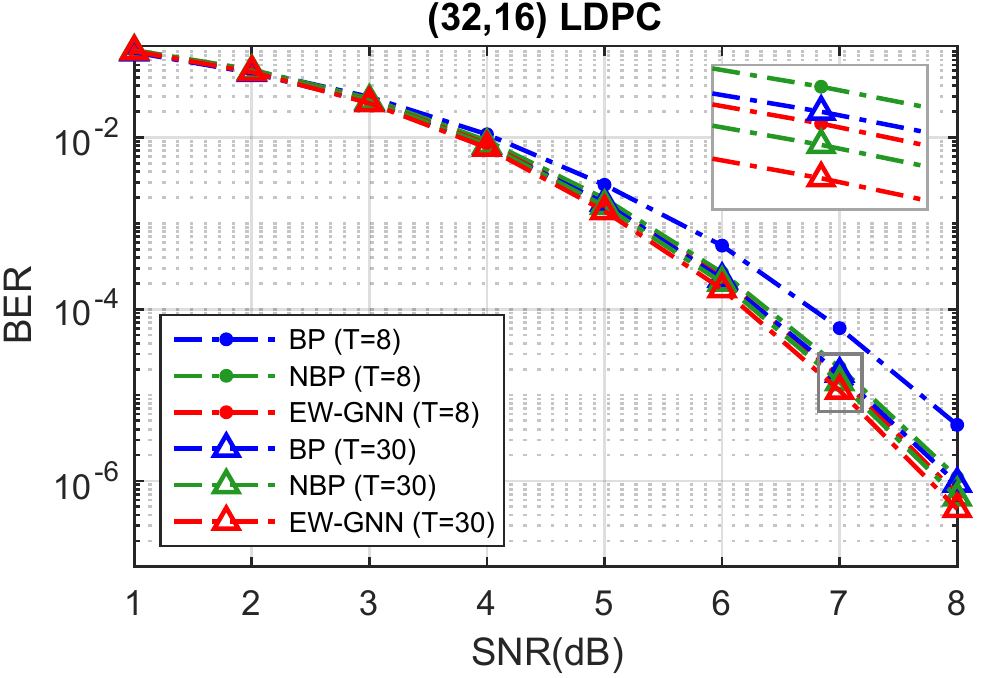}
         \caption{$n=32$}
         \label{fig::3216ldpc}
     \end{subfigure}
     \hfill
     \begin{subfigure}[b]{0.3\textwidth}
         \centering
         \includegraphics[width=\textwidth]{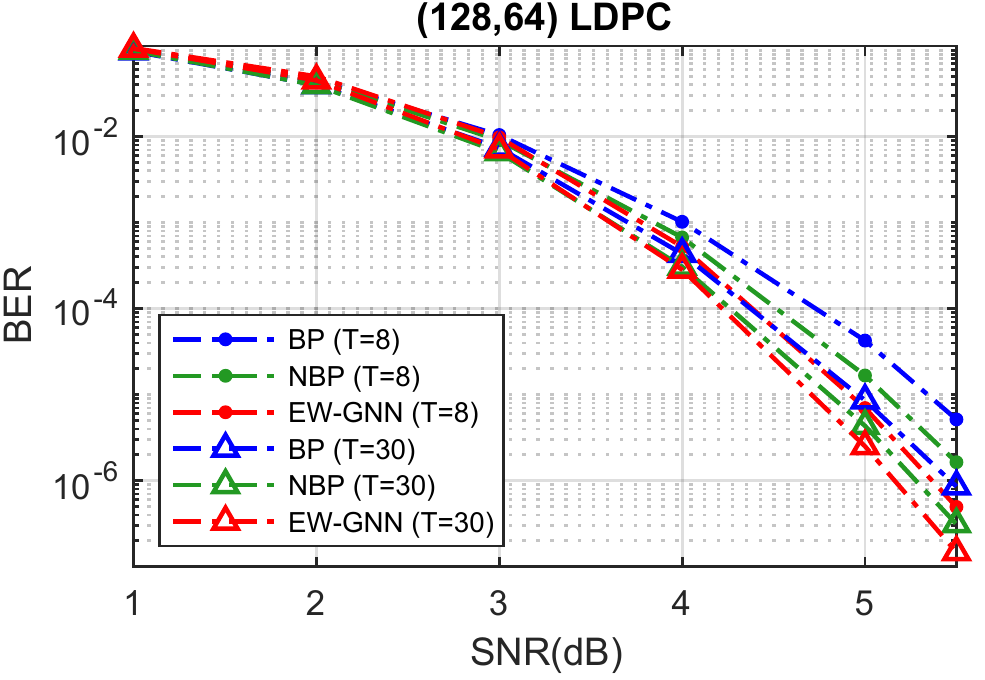}
         \caption{$n=128$}
         \label{fig::12864ldpc}
     \end{subfigure}
     \hfill
     \begin{subfigure}[b]{0.3\textwidth}
         \centering
         \includegraphics[width=\textwidth]{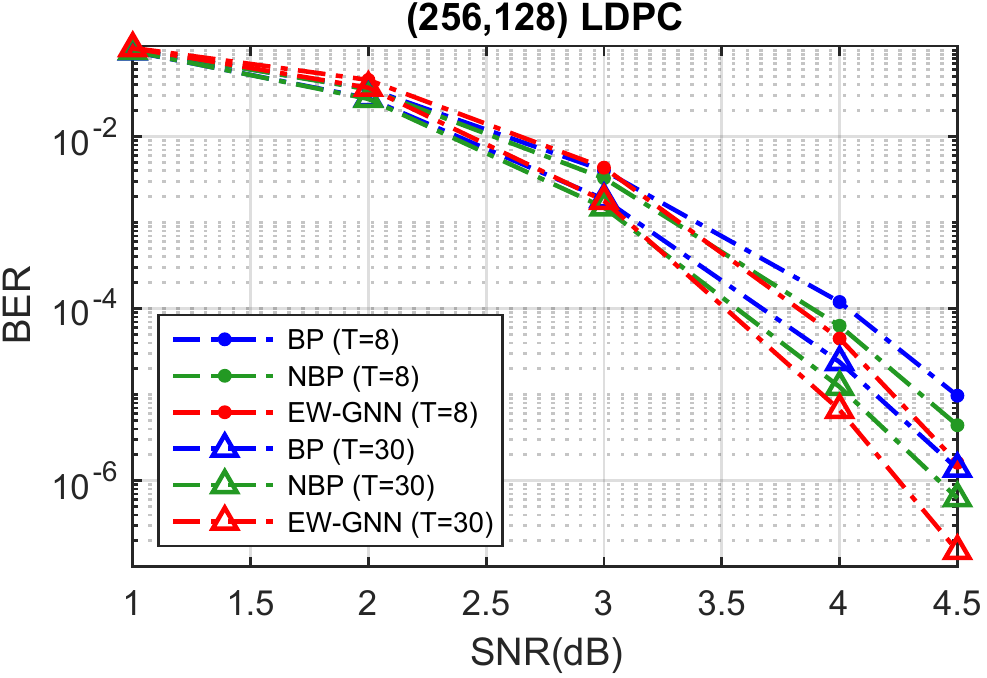}
         \caption{$n=256$}
         \label{fig::256128ldpc}
     \end{subfigure}
     \vspace{-0.5em}
        \caption{BER performance for LDPC codes at rate $1/2$. EW-GNN is only trained with the $(32,16)$ LDPC code and $T=8$.}
        \label{fig::ldpc_results}
        \vspace{-1.5em}
    \centering
\end{figure*}

\vspace{-0.5em}
\subsection{BCH Codes}
\vspace{-0.3em}
We first train an EW-GNN decoder for the $(63,51)$ BCH code with $T=8$ decoding iterations, and consider two inference scenarios with different number of decoding iterations, $T=8$ and $T=30$. 

\subsubsection{{BER performance}}
As shown in Fig. \ref{fig::6351bch},
EW-GNN outperforms BP and NBP in terms of BER for $(63,51)$ BCH code with both $T=8$ and $T=30$.  When $T=8$, our proposed EW-GNN decoder has $1.2 ~\mathrm{dB}$ and $0.62 ~\mathrm{dB}$ larger coding gains compared to the BP and NBP algorithms, respectively, at high SNRs.
Although EW-GNN is trained {in the scenario} with $T=8$, its BER performance remains superior with a larger number of decoding iterations. 
As shown, when $T=30$, the EW-GNN decoder outperforms the NBP decoder by $0.61 ~\mathrm{dB}$.
It is worth noting that our proposed decoder with $T=8$ has a better BER performance than the NBP decoder with $T=30$.

\subsubsection{Scalability}
We directly apply the EW-GNN model trained for the $(63,51)$ BCH code with $T=8$ to decode $(63,45)$ and $(63,36)$ BCH codes. 
Note that the Tanner graph of $(63,51)$ BCH codes has much fewer edges (i.e., smaller network size) than those of $(63,45)$ and $(63,36)$ BCH codes.
As can been seen in Fig{s}. \ref{fig::6345bch} and \ref{fig::6336bch}, the EW-GNN algorithm still achieves better BER performance than the BP and NBP algorithms for $(63,45)$ and $(63,36)$ BCH codes.
For example, when decoding the $(63,36)$ BCH code with $T=30$, the proposed EW-GNN increases the coding gain by $0.8 ~\mathrm{dB}$ and $0.2 ~\mathrm{dB}$ compared to BP and NBP respectively. 

\vspace{-0.5em}
\subsection{LDPC Codes}
\vspace{-0.3em}
In this subsection, we train the EW-GNN decoder for the half-rate CCSDS $(32,16)$ LDPC code \cite{ccsds2015short}  with the number of decoding iterations $T=8$.
\subsubsection{BER performance}
As depicted in Fig. \ref{fig::3216ldpc}, the proposed EW-GNN algorithm  achieves the best decoding performance among all its counterparts in decoding the $(32,16)$ LDPC code.
Specifically, when $T=30$, EW-GNN improves the coding gains of BP and NBP by $0.21 ~\mathrm{dB}$ and $0.1 ~\mathrm{dB}$ at high SNRs, respectively.

\subsubsection{Scalability}
To investigate the scalability of EW-GNN to different code lengths, the EW-GNN model trained for the $(32,16)$ LDPC codes is directly applied to longer LDPC codes at the same code rate. As shown in Fig. \ref{fig::12864ldpc}, when $n=128$, the EW-GNN decoder still outperforms the BP decoder and the NBP decoder in terms of BER. Furthermore, when $n=256$ as illustrated in Fig. \ref{fig::256128ldpc}, EW-GNN outperforms BP and NBP by $0.3 ~\mathrm{dB}$ and $0.2 ~\mathrm{dB}$ of the coding gain at {high SNRs}, respectively. Therefore, we highlight that even though EW-GNN is trained with an extremely short code ($n=32$), {it can achieve better} BER performance when applied to longer codes {without fine-tuning}. This property enables {the use of EW-GNN in practical communication systems with dynamic code lengths}. 
{In addition}, one can train the EW-GNN with very short codes to decode much longer codes, to save the training time and memory consumption. 
\vspace{-0.2em}
\section{Conclusion}
\vspace{-0.2em}
In this paper, we developed a scalable GNN-based decoding algorithm for short linear block codes, referred to as the edge-weighted graph neural network (EW-GNN) decoder. 
Simulation results show that EW-GNN can improve BER performance compared to the conventional BP and the NBP decoder for short BCH and LDPC codes. EW-GNN has training complexity independent of codeword lengths, which is lower than that of NBP for long/dense linear codes. 
Moreover, EW-GNN is scalable to the code length and code rate, i.e., a well-trained EW-GNN can be used to decode codes with different parameters without re-training. 

\vspace{-0.5em}
\bibliographystyle{ieeetr}
\bibliography{main}
\vspace{-0.5em}

\end{document}